\documentclass[20pt]{article}
\usepackage{epsfig}
\usepackage{amsfonts}
\begin {document}

\title {FROM  GEODESICS OF THE MULTIPOLE SOLUTIONS TO THE PERTURBED KEPLER PROBLEM}

\author{J.L. Hern\'andez-Pastora\thanks{E.T.S. Ingenier\'\i a
Industrial de B\'ejar. Phone: +34 923 408080 Ext
2263. Also at +34 923 294400 Ext 1527. e-mail address: jlhp@usal.es} $  ^{1,2}$ and J. Ospino$^1$ \\
\\
$^1$ Departamento de Matem\'atica Aplicada. \\ Universidad de Salamanca.  Salamanca,
Espa\~na.  \\
\\
$^2$ Instituto Universitario de F\'\i sica Fundamental y Matem\'aticas. \\ Universidad de Salamanca. Salamanca, Espa\~na}

 \maketitle

\begin{abstract}
A static and axisymmetric solution of the Einstein vacuum equations with a finite number of Relativistic Multipole Moments (RMM) is written  in MSA coordinates up to  certain order of approximation, and the structure of its metric components is explicitly shown. From the equation of equatorial geodesics we obtain the Binet equation for the orbits and it allows us to determine the  gravitational potential that leads to the equivalent classical orbital equations of the perturbed Kepler problem. The relativistic corrections to Keplerian motion are provided by the different contributions of the RMM of the source starting from the Monopole (Schwarzschild correction). In particular, the perihelion precession of the orbit is calculated in terms of the quadrupole and 2$^4$-pole moments. Since the MSA coordinates generalize the Schwarzschild coordinates, the result obtained allows  measurement of the relevance of the quadrupole moment in the first order correction to the perihelion frequency-shift.

\end{abstract}

\vskip 1cm
PACS numbers:  95.10.Eg, 04.80.Cc, 04.20.-q, 96.30.Dz

\newpage

\section{Introduction}

In a previous work \cite{cqg2} we proved that the static and axisymmetric solutions of the Einstein vacuum equations with a finite number or Relativistic Multipole Moments (RMM) \cite{geroch} can be described  by means of a function $u$ which resembles exactly the Newtonian multipole potential if we write the metric in a so-called MSA (Multipole Symmetry Adapted) system of coordinates. In fact, this is properly the definition of these coordinates, which are constructed  iteratively in terms of the Weyl coordinates.

The goodness of these coordinates arises from the fact that they allow us to generalize some theorems hold in Newtonian Gravity (NG) that establish the existence of  certain symmetry groups of the equations satisfied by the classical potential whose group-invariant solutions represent the solutions with a finite number of RMM in NG \cite{cqg1}. The existence of these systems of coordinates is proved to be related with the fact that the relativistic equations analogously admit those symmetry groups that lead to the Pure Multipole Solutions in General Relativity (GR) (those with a finite number of RMM).  In addition, the symmetry group for the case of spherical symmetry allows us to determine univocally  the specific MSA system of coordinates by means of a Cauchy problem, whose solution provides the standard Schwarzschild radial coordinate.

These characteristics make the MSA system of coordinates  become a very relevant reference to describe the gravity of stellar compact bodies whose multipole structure is known, or suitably estimated {\it a priori}, and differs slightly from the spherical symmetry. Moreover, a new feature of these coordinates reveals more relevance for the description of the Pure Multipole Solutions because they provide us with a procedure to establish measurements of high physical and astronomical interest about the behaviour of test particles orbiting into this kind of space-time.

Until now, only the spherical symmetry solution of Einstein vacuum equations has been written in MSA coordinates: this is the Schwarzschild metric in standard coordinates. But, do not we have  a nonspherical  axisymmetric solution written in these systems of coordinates, and it would be a very relevant success for the description of gravitational effects derived from deviations of the spherical symmetry. We are able to write explicitly all the metric components of a static solution characterized by any finite number of RMM, in particular the monopole, quadrupole and $2^4$-pole moments, in a system of coordinates such that the metric recovers the Schwarzschild limit when all RMM higher than monopole vanish. Hence, we are introducing a system of coordinates that generalizes the Schwarzschild standard coordinates and it can be used to study, in analogy with the spherical case, the physics of test particles in space-time slightly different from the Monopole Solution.

A very useful tool for the study of orbital motions in a classical gravitational problem is a second order differential equation called the Binet equation, from which the Kepler laws, for instance, can be deduced.  From the geodesics of a space-time we can obtain a relativistic Binet equation whose resolution allows to determine relativistic corrections to Keplerian motion like the correction to perihelion and node line precession. The calculation of the Binet equation in an MSA system of coordinates is specially suitable to describe the influence and relevance of the different RMM in such corrections. It provides the gravitational effects due to deviation from sphericity when comparison with the Schwarzschild corrections is made.

As is known,(see \cite{goldstein} and references therein) the GR theory predicts a correction of the Newtonian movement which can be interpreted by means of a classical potential type $r^{-3}$. The Schwarzschild spherically symmetric solution of Einstein vacuum equations corresponds with a perturbed Hamiltonian in the Kepler problem. In fact, the Binet equation shows directly this relation between the  orbital equation for  geodesics in a relativistic space-time and the orbits obtained for a classical Hamiltonian dynamical system.  

The question that  promptly arises is whether any other solution of the static and axisymmetric Einstein vacuum equations can be identified with certain classical system by means of a suitable perturbative potential to obtain equivalent equations for the orbits. This is one of the purposes of this work which has been accomplished through the determination of the equivalent potential associated to a perturbed Kepler problem  that leads to the same orbital equation as the one for certain geodesic  of any static and axisymmetric space-time. In this scenario  the study of geodesics of a test particle for a Pure Multipole Solution is particularly relevant, because the multipole characteristics of that solution allow  identification the contributions of its RMM within a classical description of the problem. And here is where MSA coordinates become absolutely prescriptive.

Standard techniques for the study of the equivalent classical problem can be performed. In particular, the first order perturbation theory throw interesting results about relativistic corrections  to the perihelion in Keplerian motion. As we will see, expectations arise from this study when  the role of the quadrupole, and the magnitude of its contribution, are considered, because certain discrepancies are obtained with respect to other similar calculations \cite{leo1}, \cite{leo2}. In fact, experiments for the measurement of the quadrupole moment can be outlined from the contribution of the monopole and quadrupole moments to the perihelion shift at first order.

This work is organized as follows. In section 2 a procedure to write any Pure Multipole Solution in MSA coordinates is shown and, in particular, all  the metric components of a static and axisymmetric solution with a finite number ($M_0$, $M_2$, $M_4$) of RMM are explicitly obtained. In order to do that, the MSA system of coordinates adapted to this kind of solution is previously calculated. The good behaviour of the metric and the structure of it, for a general case, is discussed.

In section 3 we calculate the geodesics of such a metric. We show that the restriction to the constant equatorial hyper-surface leads to a geodesic equation which can be written as a one-dimensional equivalent problem for an effective potential.

Section 4 is devoted to introduce  the Binet equation from its classical derivation, and we show the relativistic Binet equation for the Pure Multipole Solution. We derive the perturbative potential that identifies the calculation of geodesics for any Pure Multipole Solution with the resolution of a classical Hamiltonian perturbation of the Kepler-problem.

In section 5 we show the relevance of the quadrupole moment on the relativistic correction to the perihelion for a test particle in a perturbed Keplerian orbit. We put our attention on the effect of the quadrupole moment at the same  order as the monopole correction introduced by the Schwarzschild solution. This result suggests the possibility of measurement of the quadrupole moment through this contribution.

Finally, we discuss on the results obtained  in a conclusion section.

\section{The Pure Multipole Solutions in the MSA system of coordinates}

The static and axisymmetric solutions of the Einstein vacuum equations with a finite number of Relativistic Multipole Moments (RMM)  will be referred to  as the Pure Multipole Solutions  from now on. Some authors have devoted works on researching about those solutions \cite{varios}, \cite{mq1}.

The MSA (Multipole Symmetry Adapted) systems of coordinates were defined \cite{cqg2} as those that allow  writing the  metric component in terms of a function $u$ resembling the Newtonian gravitational  potential, or equivalently, we can say that the expansion of that metric component ($g_{00}=-1+2 u$) in power series of the inverse radial MSA-coordinate provides a multipole Thorne structure with vanishing Thorne's rests \cite{thorne}.
The MSA system of coordinates belongs to a class of ACMC  introduced by Thorne \cite{thorne}, with  a suitable choice of asymptotical behaviour  for the case of equatorial symmetry (the  odd order RMM are null).  The system of MSA coordinates $\{\hat x^{\alpha}\}=\{t,r,y\equiv \cos\hat \theta,\varphi\}$ ($\alpha=0..3$) can be  constructed iteratively in terms of the Weyl coordinates  $\{x^{\alpha}\}=\{t,R,\omega\equiv \cos\theta,\varphi\}$  as follows \cite{cqg2}:
\begin{eqnarray}
 r=R\left[1+\sum_{n=1}^{\infty}f_n(\omega)\frac{1}{R^n}\right] \nonumber \\
y=w+\sum_{n=1}^{\infty}g_n(\omega)\frac{1}{R^n} \ .
\label{trans}
\end{eqnarray}
For the purposes of this work we consider the gauge (\ref{trans}) up to certain order of approximation $O(1/R)$, and we proceed to calculate the system of coordinates associated to a Pure Multipole Solution with the set of RMM desired. In particular, the constructive method showed in \cite{cqg2} is followed here to calculate the MSA coordinates associated to  a solution only possessing the three first Multipole Moments of mass: $M_0\equiv M$, $M_2\equiv Q$ and $M_4\equiv D$. In the Appendix we show explicitly the expressions of the functions $f_n(\omega)$ and $g_n(\omega)$ up to order $O\left( 1/R^9\right)$ in the power of the inverse radial Weyl-coordinate $R$ (\ref{A1})-(\ref{A2}).
The solution managed  can be considered an exact Pure Multipole solution in the sense that it only poses those RMM, but nevertheless the components of the metric are written in terms of a power series of the inverse radial coordinate $r$ up to the order of approximation of the gauge (\ref{trans}), except for the $g_{00}$ whose analytic expression is as  follows:
\begin{equation}
 g_{00} =-1+\frac2{c^2} \left[ \sum_{n=0}^N \frac{M_{2n}}{r^{2n+1}}
P_{2n}(y)\right] , \label{g00}
\end{equation}
where $N+1$ is the number of RMM of the Pure Multipole Solution selected ($N=2$ for our case) and $P_{2n}(y)$ stand for the Legendre polynomials depending on the MSA angular variable $y=\cos\hat\theta$.

We first need to perform the inverse change of coordinates  (\ref{trans}) to write  the metric in the MSA coordinates from the general Weyl line element of  a static and axisymmetric vacuum space-time:
\begin{equation}
 ds^2= -e^{2\Psi} dt^2+e^{-2\Psi+2\gamma} (dR^2+R^2d\theta^2)+e^{-2\Psi}R^2\sin^2\theta  d\varphi^2 \ .
\label{dsweyl}
\end{equation}

The case of spherical symmetry  is especially relevant because we know explicitly the sum of the series appearing in the gauge (\ref{trans}) and the standard radial coordinate of Schwarzschild is recovered (see \cite{cqg2} for details). The MSA $(\{r,y\})_{M}$ system of coordinates for the spherical symmetry is the following:
\begin{eqnarray}
r&=&M+\frac R2 (r_++r_-) = M(x+1) \nonumber\\
y&=&\frac{R}{2M} (r_+-r_-) =y_p \ , \label{msaM}
\end{eqnarray}
where $r_{\pm}\equiv\sqrt{1\pm 2 \omega\lambda+\lambda^2}$,  $\lambda\equiv M/R$, and  $\{x,y_p\}$ are the prolate spheroidal coordinates \cite{prolate}, \cite{algomio}. The inverse relations between these coordinates are given by the following expressions:
\begin{eqnarray}
R&=&r\sqrt{1+y^2\hat \lambda^2-2\hat \lambda}\nonumber\\
\omega&=&\frac{y(1-\hat \lambda)}{\sqrt{1+y^2\hat \lambda^2-2\hat \lambda}} \ , \label{weylM}
\end{eqnarray}
where $\hat \lambda \equiv M/r$.
By performing the change of coordinates in the line element (\ref{dsweyl}) we obviously obtain the known Schwarzschild metric in standard Schwarzschild coordinates:
\begin{equation}
 ds^2= -\left(1-\frac{2M}{r}\right) dt^2+\left(1-\frac{2M}{r}\right)^{-1}
dr^2+r^2(d\hat\theta^2+\sin^2\hat\theta
d\varphi^2) \ . \label{dsweylmsa}
\end{equation}
The explicit expressions (\ref{msaM}-\ref{weylM}) are shown because we are able to recover these results from the general case of a Pure Multipole Solution by taking all RMM equal to zero except for the Monopole, as we will see in what follows.

The inversion of the series in the gauge (\ref{trans}) to obtain relations of the type $R=R(r,y)$, $\omega=\omega(r,y)$, for a general case, is a straightforward but  cumbersome calculation. In the Appendix  we show the inverse relations\footnote{It can be seen that the expansion of the expressions (\ref{weylM}) in power series of $\hat\lambda$  reproduces this results for the Monopole case.} (\ref{A3})-(\ref{A4}) corresponding to  the previously calculated MSA coordinates  associated to a Pure
 Multipole solution  only possessing the RMM $M_0$, $M_2$ and $M_4$.

The  transformation of coordinates leads to the following components of the metric:
\begin{eqnarray}
g_{rr}(\hat x)&=&F(\hat x)  \left[R_r^2+\frac{R^2}{1-\omega^2}\omega_r^2\right]_{x=x(\hat x)}\nonumber\\
g_{ry}(\hat x)&=& F(\hat x) \left[R_rR_y+\frac{R^2}{1-\omega^2}\omega_r \omega_y\right]_{x=x(\hat x)}  \nonumber\\
g_{yy}(\hat x)&=& F(\hat x) \left[R_y^2+\frac{R^2}{1-\omega^2}\omega_y^2\right]_{x=x(\hat x)}\nonumber\\
g_{tt}(\hat x)&=&-e^{-2\Psi(\hat x)} \ , \ g_{\varphi\varphi}(\hat x)=e^{-2\Psi(\hat x)} \left[R^2(1-\omega^2)\right]_{x=x(\hat x)} \ , \label{ghat}
\end{eqnarray}
where $\displaystyle{F(\hat x)\equiv e^{-2\Psi(\hat x)+2 \gamma(\hat
x)}}$, and the subindices denote the derivative with respect to that coordinate.

By using the expression (\ref{A2}) into (\ref{ghat}) we obtain  the following metric components, up to order $O\left(\hat\lambda^7\right)$ (although we only show here the first terms because the expressions are too large) in powers of the inverse MSA radial coordinate $r$ ($\hat \lambda\equiv M/r$):

\begin{eqnarray}
g_{rr}&=& 1+2 \hat \lambda+4 \hat \lambda^2+\left[8 +q(1-3 y^2) \right] \hat \lambda^3-\frac87 \left[-14 -3 M q(1-3 y^2)\right] \hat \lambda^4+\nonumber\\
&-&\frac{1}{28} \left[-896+d(2695 y^4 -2310 y^2 +231)+q(-240
+720  y^2)+\right.\nonumber\\
&+&\left.q^2(-2688 y^4 +2016 y^2 -224)\right] \hat \lambda^5+\nonumber\\
&+&\left[64+q \frac{56}{3}(1 -3 y^2)+q^2\left(\frac{76425}{154} y^4 -\frac{28935}{77}  y^2+\frac{6161}{154}\right)+\right. \nonumber\\
&+&\left.d\left(-\frac{5075}{11} y^4 +\frac{4350}{11}  y^2
 -\frac{435}{11}\right)\right] \hat \lambda^6+O(\hat\lambda^7)
\label{metrica1}
\end{eqnarray}
\begin{eqnarray}
g_{yy}&=&\frac{M^2}{1-y^2}\left[\frac{1}{\hat\lambda^2}-2 y^2 q \hat\lambda-\frac{3}{14}  q (5+7 y^2)\hat\lambda^2+\right.\nonumber\\
&+&\left.\frac{1}{35} \left[d(1715  y^4 -1785  y^2+210)+q(-42  y^2-94)+\right. \right.\nonumber\\
&+&\left.\left.q^2(-1848 y^4 +1890 y^2-182)\right] \hat\lambda^3+\right.\nonumber\\
&+&\left.\frac{1}{462} \left[d(89915  y^4-91560  y^2+10605)+q(-484 y^2-2508)+\right. \right.\nonumber\\
&+&\left.\left. q^2(
-99294 y^4+98616 y^2-9864)\right]\hat\lambda^4\right]+O(\hat\lambda^5)
\label{metrica2}
\end{eqnarray}
\begin{eqnarray}
g_{\varphi\varphi}&=&(1-y^2){M^2}\left[\frac{1}{\hat\lambda^2}+2 q(2y^2-1) \hat\lambda+\right.\nonumber\\
&+&\left. q \frac{1}{14}(87 y^2-51) \hat\lambda^2+ \left[(\frac{366}{35} y^2-\frac{46}{7}) q+\right.\right.\nonumber\\
&+&\left.\left.(-\frac{24}{5}y^4-2+\frac{54}{5} y^2) q^2+d(14y^4-21y^2+3)\right]\hat\lambda^3+\right.\nonumber\\
&+&\left.\left[ \frac{1}{21}(386y^2-250)q +\frac{1}{77}(-45y^4-362+2164y^2)q^2  +\right.\right.\nonumber\\
&+&\left. \left. d(\frac{1645}{66} y^4-\frac{580}{11} y^2+\frac{185}{22})\right]    \hat\lambda^4\right]+O(\hat\lambda^5) \qquad ,
\label{metrica3}
\end{eqnarray}
where $q\equiv Q/M^3$, $d\equiv D/M^5$ are dimensionless parameters representing the quadrupole and $2^4$-pole moments respectively.

Let us briefly analyze these results:

\vspace{3mm}

{\it i)} First, we can see that the cross term of the metric $g_{12}$ vanishes up to this order of approximation, and in fact, the preservation of the diagonal aspect of the metric  is a characteristic of the system of MSA coordinates. Supposing that the Jacobian of the coordinate transformation
 is regular and the inversion of coordinates is able, i.e., $det\equiv |r_Ry_{\omega}-r_{\omega}y_R|\neq 0$,
it can be seen that the metric component $g_{ry}$  vanishes since we can write the following expression
\begin{equation}
 g_{ry}(\hat x)=-\frac{F(\hat x)}{\left[(1-\omega^2)det^2\right]_{x=x(\hat x)}} \hat{LB}_1(r,y)_{x=x(\hat
x)} ,
\end{equation}
where $\hat{LB}_1(r,y)\equiv \eta^{ij}\nabla_i() \nabla_j()=R^2\partial_R()
\partial_R()+(1-\omega^2)\partial_{\omega}() \partial_{\omega}()$ represents the Laplace-Beltrami operator with respect to a  3-dimensional Euclidean metric (with axial symmetry) written in Weyl spherical coordinates (see \cite{cqg2} for details),  and  it was seen in \cite{cqg2} that the MSA system of coordinates fulfills the condition $\hat{LB}_1(r,y)=0$.

\vspace{3mm}

{\it ii)} Second, we can see from the expressions (\ref{metrica1})-(\ref{metrica3}) that the metric written in the MSA system of coordinates leads to  the Schwarzschild limit by considering equal to zero all RMM higher than the Monopole, as well as that the MSA system, given by expressions (\ref{A1})-(\ref{A2}), recovers the standard Schwarzschild coordinates.

\vspace{5mm}

{\it iii)} And finally, the $g_{00}[x=x(\hat x)]$ metric component results to be equal to (\ref{g00}) as claimed by that system of coordinates.

\vspace{5mm}

A general expression for the metric components of the Pure Multipole solution with a finite number of RMM ($M$, $Q$, $D$) in MSA coordinates is the following:
\begin{eqnarray}
g_{tt}(\hat x)&=&-1+2 \left[  \hat\lambda+\frac{Q}{M^3}\hat\lambda^3 P_{2}(y)+\frac{D}{M^5} P_4(y)\right] \nonumber\\
 g_{rr}(\hat x)&=&\frac{1}{1-2\hat \lambda}\left[1+(1-2\hat\lambda)\sum_{i=3}^{\infty}\hat\lambda^i
U_i(y,Q,D)\right] \nonumber\\
g_{yy}(\hat x)&=&\frac{1}{1-y^2}\frac{M^2}{\hat \lambda^2}\left[1+\sum_{i=3}^{\infty}\hat\lambda^i
D_i(y,Q,D)\right] \nonumber\\
g_{\varphi \varphi}(\hat x)&=&(1-y^2)\frac{M^2}{\hat \lambda^2}\left[1+\sum_{i=3}^{\infty}\hat\lambda^i
T_i(y,Q,D)\right] \ ,
 \label{chachis}
\end{eqnarray}
where $P_n(y)$ stands for the Legendre polynomials, and the $U_i$, $D_i$ and $T_i$ denote polynomials in the angular variable $y$ of even order depending on the higher RMM ($Q$ and $D$ for this case).

\section{The geodesics of the Pure Multipole solutions}

We proceed now to calculate the geodesics of the metric associated to a Pure Multipole solution with mass, quadrupole and $2^4$-pole moments ($M$, $Q$, $D$ respectively)  in the corresponding MSA system of coordinates.
The set of equations for the geodesics is the following
\begin{eqnarray}
 &&\frac{d^2 t}{ds^2}g_{00}+\frac{dt}{ds}\frac{dr}{ds} \partial_rg_{00}+\frac{dt}{ds}\frac{d\hat\theta}{ds}
\partial_{\hat\theta}g_{00}=0 \label{te}\\
&&2\frac{d^2 r}{ds^2}g_{11}+\left(\frac{dr}{ds}\right)^2 \partial_rg_{11}+2\frac{dr}{ds}\frac{d\hat\theta}{ds} \partial_{\hat\theta}g_{11}-\left(\frac{dt}{ds}\right)^2\partial_rg_{00}+\nonumber\\
&&-\left(\frac{d\hat\theta}{ds}\right)^2\partial_rg_{22}-\left(\frac{d\varphi}{ds}\right)^2\partial_rg_{33}=0\label{erre}\\
&&2\frac{d^2 \hat\theta}{ds^2}g_{22}+\left(\frac{d\hat\theta}{ds}\right)^2 \partial_{\hat\theta}g_{22}+2\frac{dr}{ds}\frac{d\hat\theta}{ds} \partial_{r}g_{22}-\left(\frac{dt}{ds}\right)^2\partial_{\hat\theta}g_{00}+\nonumber\\
&&-\left(\frac{d\hat\theta}{ds}\right)^2\partial_{\hat\theta}g_{22}-\left(\frac{d\varphi}{ds}\right)^2\partial_{\hat\theta}g_{33}=0\label{teta}\\
&&\frac{d^2 \varphi}{ds^2}g_{33}+\frac{d\varphi}{ds}\frac{dr}{ds} \partial_rg_{33}+\frac{d\varphi}{ds}\frac{d\hat\theta}{ds}
\partial_{\hat\theta}g_{33}=0\label{fi} \ ,
\end{eqnarray}
where $s$ denotes the affine parameter along the geodesic. Since the metric is static and axisymmetric, both Killing vectors, $\xi$ and $\eta$, representing the corresponding isometries, remain constant along the geodesics, i.e., $\xi^{\alpha}z_{\alpha}=h$, $\eta^{\alpha}z_{\alpha}=l$, $z^{\alpha}$ being the tangent vector to the geodesic,  $h$, $l$ are constants, and therefore the Eqs. (\ref{te}) and (\ref{fi}) become $\displaystyle{g_{00}\frac{dt}{ds}=h}$ and $\displaystyle{g_{33}\frac{d\varphi}{ds}=l}$ respectively. The norm of the tangent vector $z^{\beta}z_{\beta}\equiv\epsilon$ can be written as follows:
\begin{equation}
 g_{11}\left(\frac{dr}{ds}\right)^2+g_{22}\left(\frac{d\hat\theta}{ds}\right)^2=k \label{ka} \ ,
\end{equation}
with $k\equiv\displaystyle{\epsilon-\frac{h^2}{g_{00}}-\frac{l^2}{g_{33}}}$, and hence, the geodesic equations (\ref{erre}) and  (\ref{teta}) can be written as follows:
\noindent \begin{eqnarray}
&\displaystyle{\frac{d^2r}{ds^2}+\left(\frac{dr}{ds}\right)^2\partial_r
\ln\sqrt{g_{22}g_{11}}+\left(\frac{dr}{ds}\right)\left(\frac{d\hat\theta}{ds}\right)\partial_{\hat\theta}\ln g_{11}=\frac{1}{2 g_{11}}\left[\partial_rk+k\partial_r \ln g_{22}\right]}\nonumber\\
& \label{p1}\\
&\displaystyle{\frac{d^2\hat\theta}{ds^2}+\left(\frac{d\theta}{ds}\right)^2\partial_{\hat\theta}
\ln\sqrt{g_{22}g_{11}}+\left(\frac{dr}{ds}\right)\left(\frac{d\hat\theta}{ds}\right)\partial_{r}\ln g_{22}=\frac{1}{2 g_{22}}\left[\partial_{\hat \theta}k+k\partial_{\hat \theta} \ln g_{11}\right]}\nonumber\\
& \label{p2}
\end{eqnarray}

Let us consider now the case of geodesics with constant $\hat\theta$ and $\displaystyle{\frac{d\varphi}{ds}\neq 0}$, i.e., they are not radial geodesics but those constrained to a constant hypersurface ($\hat\theta=\hat\theta_0$) with coordinates $\{t,r,\varphi\}$. This restriction is compatible with the Eq. (\ref{p2}) since it leads to
\begin{equation}
 \left[\partial_{\hat\theta}k+k\partial_{\hat\theta} \ln
g_{11}\right]_{\hat\theta=\hat\theta_0}=0\label{condi} \ ,
\end{equation}
which is fulfilled at the equatorial plane ($\hat\theta_0=\pi/2$ or equivalently $y=0$) because the expressions (\ref{chachis}) allow us to hold\footnote{Let us note that we have considered equatorial symmetry and hence   $g_{00}$ only contains even order Legendre polynomials, and the polynomials $U$, and $T$ depend on even powers of $y$.} that $(\partial_yg_{00})|_{y=0}=(\partial_yg_{11})|_{y=0}=(\partial_yg_{33})|_{y=0}=0$.
Therefore, with respect to the relevant geodesic (\ref{p1}), and by using the Eq. (\ref{ka}), we obtain on the equatorial plane the following expression:
\begin{equation}
 \frac{d^2r}{ds^2}+\frac 12 \frac{k}{g_{11}} \partial_r \ln \left(\frac{g_{11}}{k}\right)=0 \ . \label{p1f}
\end{equation}
This equation can be written as follows:
\begin{equation}
 \left(\frac{dr}{ds}\right)^2+V_{eff}=C \label{ocho} \ ,
\end{equation}
$V_{eff}$ being  a so-called effective potential which can be obtained  by integration as follows
\begin{equation}
 V_{eff}=\int \frac{k}{g_{11}} \partial_r \ln\left(\frac{g_{11}}{k}\right) dr = -\frac{k}{g_{11}} \
.\label{eff}
\end{equation}
As can be seen, the Eqs. (\ref{ocho}), (\ref{eff})  reproduce the Eq. (\ref{ka}) for constant $\hat\theta$.

\section{The Binet equation}

As is known, the classical problem of Kepler consists on determining the movement of a test particle within a gravitational field generated by a potential of the type $V(r)\sim \mu/r$. Since this potential is conservative and leads to a gravitational force orientated towards the center of the source,  the conservation of the energy $E$ for a particle with mass $m$ and  orbital angular moment $\vec J$   is followed ($\mu\equiv GMm$):
\begin{eqnarray}
 &&E=\frac 12\left(m\dot r^2+r^2m\dot \theta^2+mr^2\sin^2 \theta\dot\varphi^2\right)-GM\frac mr\nonumber\\
&&\vec J=m\vec x\wedge\vec v \Rightarrow \vec x \cdot \vec J = 0 \ , \label{conser}
\end{eqnarray}
and the  particle moves on a constant plane $\theta=\pi/2$. Hence, one may define an effective potential in the following way
\begin{equation}
 E=\frac 12m\dot r^2+\Phi_{eff}(r) \qquad , \qquad \Phi_{eff}\equiv \frac{J^2}{2mr^2}-GM\frac mr \ ,
\label{esta}
\end{equation}
and the equations of motion are the following:
\begin{equation}
 \dot r^2\equiv\left(\frac{dr}{dt}\right)^2=\frac{2E}{m}-\frac{J^2}{m^2r^2}+\frac{2GM}{r} \qquad , \qquad
\dot \varphi \equiv \left(\frac{d\varphi}{dt}\right)=\frac{J}{mr^2}. \label{conser2}
\end{equation}
Consequently, the equation of the orbits is given by the following expression in terms of a variable $u\equiv 1/r$:
\begin{equation}
 \left(\frac{du}{d\varphi}\right)^2+u^2=\frac{2m}{J^2}\left(E+GMm u\right) \ , \label{bipre}
\end{equation}
and from it we can easily  obtain (by deriving with respect to $\varphi$ the Eq. (\ref{bipre})) a second order differential equation for the orbit of the test particle:
\begin{equation}
\frac{d^2u}{d\varphi^2}+u=\frac{GMm^2}{J^2} .
\label{binetcla}
\end{equation}
This is the Binet equation which can be solved to derive the three laws of Kepler concerning the orbit of a test particle describing a closed ellipse around the source, for the attractive case $E<0$.

Let us consider now a perturbation of the Newtonian potential of the type $-\alpha/r^3$. It is straightforward to see that a generalization of the Binet equation is obtained for this case as follows:
\begin{equation}
 \frac{d^2u}{d\varphi^2}+u=\frac{GMm^2}{J^2}+3 \frac{\alpha m}{J^2} u^2 .
\label{binetper}
\end{equation}
Since the corrected potential, i.e., $V(r)=\displaystyle{-\frac{GMm}{r}-\frac{\alpha}{r^3}}$ does not depend on the angular variables hence, the gravitational force is still central and conservative and we can go on considering the conservation of the orbital angular moment of the particle moving along the orbit on a constant surface.

If we calculate the geodesics of the Schwarzschild metric in standard coordinates (\ref{dsweylmsa}) we obtain, for a constant angular variable $\hat\theta=\pi/2$, the following equation:
\begin{equation}
 \left(\frac{dr}{ds}\right)^2+\left(1-\frac{2M}{r}\right)\left(\epsilon+\frac{l^2}{r^2}\right)=h^2 \quad
,\label{schwar}
\end{equation}
where $h$ and $l$ are constants, derived from the isometries as mentioned in previous section, that represent the energy and the angular moment per unit mass respectively. In fact, this Eq. (\ref{schwar}) is exactly the Eq. (\ref{ocho}) with the effective potential  for the spherical symmetry case given by the following expression:
\begin{equation}
 V_{eff}(r)=\left(1-\frac{2M}{r}\right)\left(\epsilon+\frac{l^2}{r^2}\right) \quad .\label{potM}
\end{equation}
From the Eq. (\ref{schwar})  the orbit of the particle is described by the following equation
\begin{equation}
\frac{d^2u}{d\varphi^2}+u=-\epsilon\frac{M}{l^2}+ 3 M u^2 \ ,\label{binetrela}
\end{equation}
$u\equiv1/r$ being the inverse of the standard-Schwarzschild radial coordinate. Since $l$ is considered to  be the angular moment of the particle per unit mass ($l=J/m$) we can hold by comparing this equation with (\ref{binetper}) the following statement \cite{goldstein}:

{\it The orbit on  a constant surface $\hat\theta=\pi/2$ corresponding to a timelike geodesic ($\epsilon=-1$) of the Schwarzschild space-time  is given by the same equation as the Binet equation corresponding to the Newtonian Kepler problem with a perturbed potential of the type $-\alpha/r^3$ with $\alpha=J^2 M/m$}.

\vspace{2mm}

As is known \cite{leo1}, \cite{leo2}, the {\it relativistic} Binet equation (\ref{binetrela}) allow us to calculate corrections to the Newtonian orbit even for the motion around a spherical distribution of mass, which is no longer closed. In particular this relativistic effect has been tested in our  solar system  and it amounts  to a slow precession of the perihelion of the orbit of Mercury. Predictions of the General Relativity correcting the classical gravity, as the precession of the perihelion or deflection of light,  arises from the resolution of Binet equation and constitute the well known ests of GR.

\subsection{The generalized Binet equation}

We want to extend the above-mentioned statement to any static and axisymmetric space-time corresponding to  a nonspherical mass distribution, by means of a  generalized relativistic Binet equation. Of course that   RMM higher than the mass  will also contribute to the precession effect and, in principle,  these moments could be calculated  by measuring the precession  of a certain number of  test particles  orbiting at conveniently different distances from the gravitational source. We will devote  last section of this work to discuss  this point.

Therefore, we need to compare the relativistic Binet equation with a classical Binet equation corresponding to certain gravitational potential. According  to Eq. (\ref{ocho}) we can define an effective potential in such a way that the orbital equations (on  the equatorial plane) corresponding to geodesics of any static an axisymmetric space-times are exactly given by the classical Binet equation related to some perturbed Kepler problem.

Hence, we have established a relationship between the equation for geodesics in a relativistic space-time and the orbital equations  for an associated Newtonian potential. But moreover, since we have constructed Pure Multipole solutions with a meaningful physical interpretation in terms of their finite number of RMM as successive corrections to the spherical symmetry, the Binet equation for these solutions will provide us with the contributions of the different RMM to the relativistic effects correcting the Newtonian orbits. In order to support this assertion, we must  remind that the MSA system of coordinates used to write the metric and its geodesics are adapted to the set of RMM of the solution,  and these coordinates recover the standard Schwarzschild coordinates in the limit $M_{2n}=0$, $\forall n>0$.

In fact, from Eqs. (\ref{ocho}) and (\ref{eff}) we have that
\begin{equation}
 \left(\frac{du}{d\varphi}\right)^2=\frac{k}{g_{11}}\frac{g_{33}^2}{l^2} u^4 \ ,
\end{equation}
where the notation $u\equiv 1/r$ is used again. Finally, the  derivative  of the above equation with respect to the azimuthal angle $\varphi$ leads to the following equation:
\begin{equation}
 \frac{d^2 u}{d\varphi^2}=\frac 12
\frac{d}{du}\left[\frac{g_{33}}{g_{11}}u^4\left[\left(\epsilon-\frac{h^2}{g_{00}}\right)\frac{g_{33}}{l^2}-1\right]\right]  \ . \label{binetrelacomplet}
\end{equation}
For the case of a Pure Multipole solution with a finite number of RMM ($M$, $Q$ and $D$) this equation provides the generalized relativistic Binet equation,  for the orbit on the equatorial plane ($y=0$) of a test particle moving along a timelike geodesic ($\epsilon=-1$), as follows:
\begin{eqnarray}
 \frac{d^2
u}{d\varphi^2}+u&=&\frac{M}{l^2}+\left[ 3M +\frac {3Q}{2 l^2}  (5+6 h^2) \right]u^2+\frac {6}{7 l^2} M Q (-3+25 h^2) u^3\nonumber\\
&+&\left[-\frac{30}{M l^2}Q^2(1+h^2)-\frac{15}{56 l^2} 4Q(-7 l^2+2M^2(3-22 h^2))+\right.\nonumber\\
&+&\left.D(133
 +140 h^2) \right]u^4+O(u^5) \quad ,\label{miembrodcha}
\end{eqnarray}
where terms in powers of $u$ higher than $5$ have been neglected. As can be seen, if we take all RMM higher than monopole equal to zero  the relativistic Binet equation (\ref{binetrela}) is recovered.

Finally, we want to obtain explicitly which is the perturbed Kepler problem associated to this relativistic Binet equation. As we already saw, a perturbative potential of the type $-\alpha/r^3$ leads to a classical  orbital equation (\ref{binetper}), that  reproduces exactly the orbital equation corresponding to  a timelike geodesic of the Schwarzschild metric on the equatorial plane. In analogy with this result, if we handle with a Pure Multipole solution with a finite number of RMM, a suitable perturbed potential to the Kepler problem can be selected to obtain a generalized Binet equation that equals the Eq. (\ref{miembrodcha}). Nevertheless,  despite  the spherical symmetry case, we will see now that this potential  is given up to a suitable order of approximation in the power of the variable $u$.

Accordingly to Eq. (\ref{esta}) let us consider now a generalized  effective classical potential $\Phi_{eff}(r)$ given by this expression:
\begin{equation}
\Phi_{eff}\equiv \frac{J^2}{2mr^2}-GM\frac mr +V_p(r) \qquad , \qquad V_p(r)=-\frac{\alpha}{r^3}+V_p^{RMM}(r)\ ,
\end{equation}
where $V_p^{RMM}(r)$ denotes  the new perturbation considered in addition to the previously studied. The comparison of the resulting Binet equation with the expression (\ref{miembrodcha}) will supply us with  the perturbed potential that provides different contributions due to the RMM higher than monopole. The conservation of energy $E$ and the orbital angular momentum of the particle $J$ lead to the following equation for the orbit\footnote{Let us note that (\ref{defi}) can be obtained from Eqs.  (\ref{conser2})}:
\begin{equation}
 d\varphi=\frac{J/r^2 m}{\sqrt{2/m(E-\Phi_{eff})}}dr \ ,\label{defi}
\end{equation}
or equivalently, we can write the following generalization of the Eq. (\ref{bipre}):
\begin{equation}
 \left(\frac{du}{d\varphi}\right)^2+u^2=\frac{2m}{J^2}\left(E+GMm u+\alpha u^3-V_p^{RMM}(r)\right) \ ,
\label{bipregen}
\end{equation}
and therefore, the generalized classical Binet equation is
\begin{equation}
 \frac{d^2u}{d\varphi^2}+u=\frac{GMm^2}{J^2}+3 \frac{\alpha m}{J^2}
u^2-\frac{m}{J^2}\frac{d}{du}V_p^{RMM}(r) \ .
\label{binetclagen}
\end{equation}

By comparing this Eq. (\ref{binetclagen}) with (\ref{miembrodcha}) we can hold the following statement:

{\bf Proposition:}

{\it The classical Kepler problem for a mass distribution $M$, with total energy $E$ and orbital angular momentum $J$, perturbed with a potential given by this expression:}
\begin{eqnarray}
 V_p(r)&=&-\alpha u^3-Q m G (\frac 52+3 h^2) u^3+Q M m
G(\frac{9}{14}-\frac{75}{14} h^2)u^4+\nonumber\\
&+&\left[-\frac {3}{2m} Q J^2-6 \frac mM G (1+h^2) Q^2
 +(\frac{57}{8}+\frac{15}{2} h^2) m G D+\right.\nonumber\\
&+&\left.(\frac 97 -\frac{66}{7} h^2)m M^2 G
Q\right]u^5+O(u^6)
\nonumber\\
\label{vpr}
\end{eqnarray}
{\it provides (up to the desired order in the power of the variable $u$) the same orbital equation for a test particle of mass $m$,  as the corresponding orbital equation for the timelike geodesic on the equatorial plane of the relativistic Pure Multipole solution with a finite number of RMM if the following values of the parameters are considered}:
\begin{equation}
l=\frac{J}{mG^{1/2}} \qquad , \qquad h^2=-\frac{2E}{mG}-1 \qquad , \qquad \alpha=\frac{M J^2}{m} \ .
\label{vpr2}
\end{equation}

Let us note that the above expressed  potential can be written in terms of the dimensionless parameter $\hat\lambda\equiv M u=M/r$, which controls the perturbational character of it, as follows:
\begin{eqnarray}
& &V_p^{RMM}(r)=-mG\left(\hat\lambda^3 \frac q2 (5+6h^2)+\frac{3}{14} \hat\lambda^4 q (-3+25h^2)+\right.\nonumber\\
&+&\left.\frac{3}{14} \hat\lambda^5\left[q\left(44 h^2-6+7\frac{J^2}{M^2m^2G}\right)+28q^2(1+h^2)-\frac 74 d(19+20h^2)\right]\right) ,\nonumber\\
\end{eqnarray}
where $q\equiv Q/M^3$ and $d\equiv D/M^5$ denote for dimensionless parameters associated to the quadrupole $Q$ and $2^4-$pole moment $D$ respectively.

Consequently, the  Binet equation is given by the following expression:
\begin{eqnarray}
&&\frac{d^2u}{d\varphi^2}+u=M \left(\frac{m}{J}\right)^2 G+\left[(\frac{15}{2}+9h^2) \left(\frac{m}{J}\right)^2 G Q+3 M\right]u^2+\nonumber\\
&+&\left[M Q G \left(\frac{m}{J}\right)^2\left(-\frac{18}{7}+\frac{150}{7}h^2\right)
\right]u^3+\nonumber\\
&+&\left[\frac{15}{2} Q+\left(\frac{m}{J}\right)^2G\left(30  (1+h^2) \frac{Q^2}{M}
+M^2 Q \frac{15}{7}(-3+22 h^2)+\right.\right.\nonumber\\
&+&\left.\left.\frac{15}{8}D(20h^2+19)\right)\right]u^4 \ .
\end{eqnarray}

\section{Measuring the quadrupole moment}

The Eq. (\ref{vpr}) shows that a quadrupole contribution arises at order $u^3$, the same order as the known Schwarzschild correction due to the monopole. The role of the quadrupole becomes relevant enough when the relativistic correction to the perihelion of a test particle orbit is calculated. Hence, let us consider the above potential (\ref{vpr}) and let us develop up to first order of perturbation theory  the following perturbed kepler problem (for $n=3$):
\begin{equation}
V=-GMm\frac1r-\frac{\xi}{r^n} \quad , \quad H=H_0+\xi H_1 \quad , \quad H_1\equiv -\frac{1}{r^n} ,
\end{equation}
where $\xi$ is a small parameter, and $H_1$ is the perturbation of the  Hamiltonian $H$.

According to the standard theory of perturbations \cite{goldstein} within angular-action variables, the averaged rate of secular precession due to the perturbation is given by the following expression:
\begin{equation}
{\bar{\dot \varsigma}}=\frac{1}{\tau}\int_0^{\tau}\frac{\partial H_1}{\partial J}dt=\frac{\partial}{\partial J} \bar H_1 ,
\end{equation}
$\tau$ being the average time interval considered, in particular the period of the nonperturbed orbit; $\varsigma$ stands for the angular position of the periastron on the orbit plane, and $\bar H_1$ is the time-averaged perturbed Hamiltonian which can be calculated by using the conservation of the angular moment ($J=mr^2\dot \varphi$=cte) as follows:
\begin{equation}
{\bar H_1}=-\frac{1}{\tau}\int_0^{\tau}\frac{1}{r^n}dt=-\frac{1}{\tau}\frac{m}{J}\int_0^{2\pi}\frac{1}{r^{n-2}}d\varphi ,
\end{equation}
where $r$ can be expressed in terms of $\varphi$ by means of the nonperturbed orbit:
\begin{equation}
\frac1r=\frac 1p(1+e \cos\varphi) ,
\end{equation}
$e$ being the eccentricity and $\displaystyle{p\equiv\frac{J^2}{m^2MG}}$ is the so-called ellipse's parameter.

Therefore, the averaged rate of precession is given (for the $n=3$ case) by
\begin{equation}
{\bar{\dot \varsigma}}=\frac{6 \pi}{\tau}GM\frac{m^3}{J^4} \xi=\frac{6 \pi}{\tau}\left[GM^2\frac{m^2}{J^2}+QG^2 M\frac{m^4}{J^4}\left(\frac 52+3h^2\right)\right] \  .
\label{aver}
\end{equation}

The first term of (\ref{aver}) is the monopole contribution already known, which is provided by the Schwarzschild solution, whereas the second term is the correction to the precession due to the quadrupole moment. If we take into account the value of $h$ (\ref{vpr2}) and consider $J$ and $E$ in terms of the eccentricity $e$ and the semimajor axis $a$ of the orbital ellipse, i.e., $J^2=a(1-e^2)GMm^2$, $|E|=GMm/(2a)$, we have that
\begin{equation}
\bar{\dot \varsigma}=\frac{6\pi}{\tau}\left[\zeta+\zeta^2 q\left(-\frac 12+3\frac Ma\right)\right] \ ,
\label{averdef}
\end{equation}
where $\displaystyle{\zeta\equiv\frac{M}{a(1-e^2)}}$ is a dimensionless parameter\footnote{Let us note that $M$ must be considered in length units ($M\rightsquigarrow\frac{GM}{c^2}$). For the case of the Sun the value of this {\it length} is about $1.476$ km, and hence $M<<a$ for any planet of the solar system.} less than $1$, the quadrupole contribution is of order $\zeta^2$ and its sign depends on the relative value between $M$ and $a$:
\begin{equation}
\zeta^2 q\left(-\frac 12+3\frac Ma\right) \left\{\begin{array}{ccc}
<0&\leftrightarrow&M<<a\\
\geq 0&\Leftrightarrow&6M>a .
\end{array}
\right.
\end{equation}

In \cite{leo1}, \cite{leo2} the authors look for patterns of regularity in the sign of the contributions for different RMM, ending up  the conclusion that the linear quadrupole term is negative for a positive quadrupole. This pattern is verified by our calculation except for the case that $6M>a$, i.e., a test particle closely orbiting around a strong gravitational  source. The discrepancy arises because  in \cite{leo2} the authors do not obtain\footnote{Let us note that our parameter $\zeta$ is exactly equal to the parameter $\epsilon^2$ used in \cite{leo2} to develop the expansion series.}, up to this order in the parameter $\zeta$, the relativistic contribution $\displaystyle{\frac{3M}{a}}$ appearing in (\ref{averdef}) and the contribution of the quadrupole at order $\zeta^2$ is given only by the Newtonian term $-1/2$.

Since the first contribution of the quadrupole moment to the perihelion shift appears at the same order in $\hat\lambda$ as the monopole contribution, this calculation provides a procedure for measurement of the quadrupole moment. In principle a test particle conveniently far away ($\hat\lambda<1$) from the source should be affected  only by  the monopole and quadrupole contributions which are the only effective corrections at that distance (contributions higher than $\hat\lambda^3$, in the first order of perturbation theory, are considered negligible). Dropping any other external effects, for a suitable isolated source-particle system acting on the perihelion precession of the orbit, the quadrupole contribution can be estimated to obtain a tentative measurement of the quadrupole moment since the monopole correction is well known. In fact, the relative value between both corrections at this order of perturbation theory is the following:
\begin{equation}
\Delta\equiv\frac{Q^{term}}{M^{term}}=\frac{\zeta^2q\left(3\frac Ma-\frac12\right)}{\zeta}=\zeta q \left(3\frac Ma-\frac12\right) \quad . \label{delta}
\end{equation}

Let us calculate an estimate of this relative value for the orbit of Mercury. As is known,  Einstein's theory of gravity leads to a relative correction of the Newtonian value for  Mercury's secular perihelion drift of $42.98^{\prime\prime}/$century \cite{soffel}. The rate of precession for the relativistic monopole (Schwarzschild correction) can be calculated from the first term of our Eq. (\ref{averdef}) to obtain\footnote{The quantity $(6\pi/\tau) \zeta$ provides  an estimate of the Schwarzschild correction, with the following values of the parameters: eccentricity of the orbit $e=0.2056$, the semi-major axis of the orbit $a=6.4529 \times 10^{7} km$, $M$ being a half of the radius of Schwarzschild for the Sun $M\equiv GM_{\odot}/c^2$, and period of the Mercury's orbit $\tau=7600428 s$.} $38.01^{\prime \prime}/$century.

The quadrupole contribution can be calculated from the second term of the Eq. (\ref{averdef}), and hence, the key point of its estimate, as well as that of the relative value (\ref{delta}), depends on the dimensionless parameter $q$ related with the quadrupole moment of the Sun. This is a matter of increasing research\footnote{The gravitational multipole moments describe deviations from  a purely spherical mass distribution. Thus, their precise determination gives indications on the solar internal structure. It is difficult to compute these parameters and the way to derive the best values and how they will be determined in a near future by means of space experiments is the aim of several works \cite{rozelotpiro}.}, because the quadrupole moment is a relevant  quantity for a lot of  related measurements; in fact, the evaluation of the solar quadrupole moment still faces some controversy: on one side, the theoretical values strongly depend on the solar model used, whereas accurate measurements are very difficult to obtain from observations, in particular the value of the quadrupole moment can be inferred to be in agreement with the experiment observations of precisely the perihelion advance of Mercury (see \cite{pirozelot} and references therein).

Different estimates of the quadrupole moment of the Sun \cite{pirozelot}, \cite{rozelotpiro}, provide a range of values from a theoretical determination $J_2=(2\pm 0.4)\times 10^{-7}$ to another bounds around $J_2=(1.4 \pm1.5)\times 10^{-6}$ \cite{soffel}; nevertheless, the quadrupole moment of the Sun may not exceed the critical value of $J_2=3.0 \times 10^{-6}$ according to the argument given in \cite{rozbo}, based upon the accurate knowledge of the Moon's physical librations (these data reach accuracies at the milliarcsecond level).

The factor $M/a$ for the Mercury's orbit is about $2.257\times 10^{-8}$, and the parameter $\zeta$ strongly depends on that value, since $1/(1-e^2)$ is around $1.04415$, and so $\zeta\simeq2.356\times 10^{-8}$. Therefore,  Eq. (\ref{delta}) for the relative value of the quadrupole contribution with respect to the Schwarzschild correction leads to the following estimate: 
\begin{equation}
 |\Delta|\simeq  1.18 \times 10^{-8} q \label{deltaest} \ ,
\end{equation}
or equivalently, the rate of precession due to the quadrupole contribution is estimated as follows:
\begin{equation}\displaystyle{\bar{\dot \varsigma}_Q=\frac{6\pi}{\tau}\left[\zeta^2 q\left(-\frac 12+3\frac Ma\right)\right]}\simeq 4.48 \times 10^{-7}q{}^{\prime \prime}/century. \label{myestimate}
\end{equation}

This result 
is perfectly in agreement with the current predictions of measurements and values of the advance of Mercury's perihelion deduced from observational data: As is known \cite{pirozelot}, once correcting for the perturbation due to the general precession of the equinoxes and for the perturbation due to other planets, the advance of the perihelion of Mercury is a combination of a purely relativistic effect and a contribution from  the Sun's quadrupole moment. One may compute the corrective factor to the prediction due to General Relativity ($42.98^{\prime \prime}/$century, the Schwarzschild contribution, which does not include the quadrupole correction): this factor, i.e., the relative value of the solar quadrupole correction  with respect to the Schwarzschild contribution is 
\begin{equation}
 2.821 \times 10^{-4} (2.0\pm 0.4) \label{deltadato} \ ,
\end{equation}
 for a value of the quadrupole moment $J_2=2.0\times 10^{-7}$, or equivalently, the correction to the 
perihelion precession due to the quadrupole moment is about
\begin{equation}
 2.425 \times 10^{-2}{ }^{\prime \prime}/century \label{quadrudato} \ ,
\end{equation}
 These calculations are done within 
the Parameterized Post-Newtonian (P.P.N.) formalism describing a fully conservative relativistic theory of gravitation (see Eq. (1) in \cite{pirozelot}), and they are in accordance with our Eqs. (\ref{deltaest}) and  (\ref{myestimate}) respectively by taking into account the appropriated relation between the parameter $J_2$ in P.P.N. formalism and the relativistic multipole moment $Q$ defined by Geroch \cite{geroch}, i.e., $|Q|\equiv M R_{\odot}^2 J_2$. Hence, from (\ref{deltaest}) we have that 
\begin{equation}
 |\Delta|\simeq 4.45 \times 10^3 J_2 \simeq 8.9 \times 10^{-4} \ ,
\end{equation}
 where a value of $J_2=2.0 \times 10^{-7}$ has been used, or equivalently, Eq. (\ref{myestimate}) leads 
to the predicted relations between observational or 
theoretical quadrupole moment and the quadrupole correction to the perihelion precession of Mercury's orbit: \begin{equation}
\displaystyle{\bar{\dot \varsigma}_Q\simeq   4.48 \times 10^{-7} J_2 \left(\frac{R_{\odot}}{M}\right)^2} \simeq 1.12 \times  10^{5} J_2 {}^{\prime \prime}/century \simeq 2.24 \times 10^{-2} {}^{\prime \prime}/century.
\end{equation}

\section{Conclusion}

In this work  we have tried to show the relevance of the MSA system of coordinates for the description of static  and axisymmetric vacuum solutions with a finite number of RMM, particularly  for those which are  slightly different with respect to the spherically symmetric solution. In \cite{herrera} the authors study  the behaviour of different geodesics of quasispherical spacetime,  for example the case of self-gravitating  sources with the exterior gravitational field of the $M-Q^{(1)}$ solution \cite{mq1}.

First, we have explicitly written a static and axisymmetric vacuum solution with a finite number ($M$, $Q$, $D$) of RMM in MSA coordinates. The expressions obtained for the metric components are approximated because the MSA system of coordinates are constructed iteratively by means of a power series. But, these results allow us to handle with the Pure Multipole solutions as generalizations of the Schwarzschild solution  in the sense that each one of the metric components are written as a series  whose first term represents the monopole solution and the following terms provide the successive corrections  to the spherical symmetry due to the other multipoles (\ref{chachis}). Furthermore, the $g_{00}$ metric component is calculated to any order and it resembles, as it was desired,  the formal expression of the classical multipole potential. Since these solutions are static, we are actually giving the Ernst potential \cite{ernst} of the solution in such a way.
Until now, no other solution has been written in MSA coordinates except for the Schwarzschild solution. The expressions obtained for the metric are supported by the calculation of the MSA coordinates in terms  of the Weyl coordinates (as well as the inverse relations). We are providing a system of coordinates that generalizes the standard Schwarzschild coordinates for the cases of Pure Multipole solutions.

Second, we have used this system of coordinates to study the behaviour of test particles orbiting in a space-time described by a Pure Multipole solution. Two results seems to be specially relevant:
One of them  is the possibility of finding an equivalent classical problem; we are able to write the orbital equation associated to geodesics in the equatorial plane identically equivalent to the Binet equation obtained from a classical perturbed Kepler problem. In other words, we can describe the orbital motion of a test particle in the same way as it is studied the classical field equations for a problem of a conservative potential endowed with certain perturbation.
We calculate explicitly this perturbative potential in terms of the physical parameters of the virtual Hamiltonian dynamical system (energy $E$, particle mass $m$, orbital angular momentum $J$...) and the RMM of the vacuum solution.

This relationship between the resolution of a classical dynamical system and a geometrical description of the problem by means of an associated riemannian metric is an analogous result  to the prescription derived from the Maupertuis-Jacobi principle \cite{maupertuis}: the trajectories of a mechanical system, with a natural Lagrangian function $L=(1/2) g_{ij} \dot q^i \dot q^j -V(q)$, are geodesics of the Jacobi metric $g^J_{ij}=2(E-V)g_{ij}$ for a fixed total energy $E$ of the system.  Possibly be some connection between those schemes can be explored.

The other important result of this study of the geodesics in MSA coordinates is the calculation of high order relativistic corrections to Keplerian motion. In particular, we have calculated the perihelion shift due to the correction of the quadrupole moment in addition to the Schwarzschild contribution, and it is shown that both corrections arises at the same order in the perturbed potential. Classical techniques of perturbation theory can be used to make this calculation since the description of the relativistic motion is made from the equivalent Newtonian problem. 
In \cite{leo1}, \cite{leo2} the authors develop a calculation of relativistic corrections to Keplerian motion; the advantage of our work is that we do not need to introduce a parameter {\it ad hoc} to perform the expansion series, because the MSA system of coordinates itself leads to geodesics of the solution written in such a way that the corrections due to any RMM are clearly distinguished.

The estimate of the quadrupole  contribution leads to the conclusion that it is small compared  with the Schwarzschild correction for the case of Mercury's orbit, but it is perfectly  detectable by present experiments, since one of the techniques used for measuring the quadrupole moment of the Sun is just by means of the perihelion precession of the orbit \cite{rozelotpiro}, \cite{pirozelot}.

Comparison  between both corrections (\ref{delta}) leads to the conclusion that the quadrupole contribution to the perihelion precession may be quantitatively significant with respect to the Schwarzschild correction for scenarios with a very massive source and test particles closely orbiting around: Eq. (\ref{averdef}) shows that it is important not only the value of the quadrupole moment but also the factor $M/a$. So, if we think about an experiment or astrophysical system with a test particle closely orbiting around a strong gravitational source we could handle with a value of the factor $M/a$ which may compensate the  factor $q$ whose value is rather small for standard astrophysical systems (except for compact binary systems).

A future generalization of our results to a stationary and axisymmetric case will provide us with a more realistic scenarios, and the estimates will become very relevant if the calculation is applied for instance to an axisymmetric spinning star.
For example, neutron stars are extremely compact objects typically around $1 M_{\odot}$ or $2 M_{\odot}$ compressed in a radius of a few kilometers, and hence, the corresponding value of $M/a$ for a test particle orbiting closely around a neutron star  can be near to $10^{-1}$.

In addition, some works have studied the central role of the innermost stable circular orbit (ISCO) in the relativistic precession of orbits around neutron stars \cite{stergi}, \cite{morsink}. Strange stars has also been considered as relevant sources  (of QPOs for example \cite{gourgou}); Strange stars are objects with two main characteristics: they are made of a mixture of quarks and they have no minimum mass. It has also been shown in \cite{stergi},\cite{gourgou} that ISCO is located outside the strange star for relatively high mass (ranged from 1.4 to 1.6 solar masses depending on the different EoS) even at very high  rotation rates,  this fact being the main difference with respect to neutron stars.
In \cite{ss} Shibata and Sasaki calculated analytically  the ISCO of neutral  test particles  around a massive, rotating and deformed source in vacuum, including the first four gravitational multipole moments (following the scheme given  by Shibata the authors in \cite{sanabria} extend it to the electrovacuum case.); it is possible to study the role of the quadrupole moment of mass related to the neutron stars, whose value is not constrained to be small, and the factor $M/a$ can be considered for hypothetical orbits nearby the ISCO ($r\approx 6M$).

\section{Appendix}

The gauge (\ref{trans}) is defined in terms of series, up to order $O(1/R^9)$, with the following functions $f_n(\omega)$ and $g_n(\omega)$:

\begin{eqnarray}
f_1(\omega)&=&M \quad , \quad
f_2(\omega)=\frac 12 M^2 (1-\omega^2)\quad , \quad
f_3(\omega)=\frac 12 Q(1-3 \omega^2)\nonumber\\
f_4(\omega)&=&M Q\left(-\frac 54  \omega^4-\frac{19}{28} +\frac{39}{14} \omega^2\right)+M^4\left(\frac 34 \omega^2-\frac 58 \omega^4-\frac 18\right)\nonumber \\
f_5(\omega)&=& \frac{Q^2}{M}(1+12 \omega^4-9 \omega^2)+M^2 Q \omega^2(2-3  \omega^2)+D\left(\frac{45}{4} \omega^2-\frac{105}{8} \omega^4-\frac{9}{8}\right)\nonumber\\
f_6(\omega)&=&Q M^3\left(\frac{73}{168}+\frac{705}{56}  \omega^4-\frac{361}{56} \omega^2-\frac{45}{8}  \omega^6\right) +\nonumber\\
&+&M^6\left(\frac{35}{16} \omega^4-\frac{15}{16} \omega^2-\frac{21}{16} \omega^6+\frac{1}{16}\right)+\nonumber\\
&+&M D\left(\frac{191}{88}-\frac{1965}{88} \omega^2-\frac{21}{8} \omega^6+\frac{2485}{88}  \omega^4\right) +\nonumber\\
&+&Q^2\left(-\frac{1247}{616} +\frac{10347}{616}  \omega^2-\frac{13089}{616} \omega^4-\frac{15}{8} \omega^6\right)\nonumber\\
f_7(\omega)&=& Q^2 M\left(\frac{61}{28}-\frac{2711}{140}\omega^4-\frac{38}{5}  \omega^2+\frac{81}{2} \omega^6\right)+\nonumber\\
&+&M^4 Q\left(-2 \omega^2
+\frac{17}{2}  \omega^4-\frac{15}{2}  \omega^6\right)+\nonumber\\
&+& M^2 D \left(\frac32 \omega^2+\frac{85}{2}  \omega^4-\frac 32-\frac{105}{2}  \omega^6 \right)+\nonumber\\
&+&\frac{Q D}{M}\left(-3+154 \omega^6-195 \omega^4
+60  \omega^2\right)+\nonumber\\
&+&\frac{Q^3}{M^2}\left(
-\frac{144}{5} \omega^2+\frac{981}{10}  \omega^4-\frac{414}{5} \omega^6
+\frac 32\right)\nonumber\\
f_8(\omega)&=&\frac{Q^3}{M}\left(\frac{52341}{520} \omega^2-\frac{138851}{24024}-\frac{13553439}{40040} \omega^4+\frac{892359}{3080} \omega^6\right)+\nonumber\\
&+&D Q\left(-\frac{84665}{176}\omega^6+\frac{646781}{64064}+\frac{19850195}{32032} \omega^4-\frac{280173}{1456} \omega^2-\frac{675}{64} \omega^8\right)+\nonumber\\
&+&Q M^5 \left(
-\frac{383}{1056} -\frac{4897}{112} \omega^4+\frac{4825}{462} \omega^2+\frac{1595}{28}  \omega^6-\frac{715}{32}  \omega^8\right)+\nonumber\\
&+&M^8\left(-\frac{5}{128}-\frac{315}{64} \omega^4+\frac{35}{32}  \omega^2
+\frac{231}{32}  \omega^6-\frac{429}{128} \omega^8\right)+\nonumber\\
 &+&M^3 D\left(\frac{445}{2288}+\frac{3890}{143}  \omega^2+\frac{3171}{22} \omega^6-\frac{159915}{1144}  \omega^4-\frac{273}{16}  \omega^8\right)+ \nonumber\\
 &+&M^2 Q^2\left(-\frac{936731}{672672}-\frac{1143841}{140140}  \omega^2
+\frac{34285679}{560560}  \omega^4-\frac{4121}{77}  \omega^6-\frac{715}{32}  \omega^8\right)\nonumber\\
\label{A1}
\end{eqnarray}

\begin{eqnarray}
g_1(\omega)&=&0 \quad , \quad g_2(\omega)=(1-\omega^2) (-\frac 12 \omega M^2)\nonumber\\
g_3(\omega)&=&(1-\omega^2) (-Q \omega)\nonumber\\
g_4(\omega)&=&(1-\omega^2) \omega(-\frac{1}{56})\left[ M^4 (49 \omega^2-21)+ Q(70 \omega^2-78) M \right]\nonumber\\
g_5(\omega)&=&(1-\omega^2) \omega \frac{1}{10 M} \left[D M(-105 \omega^2+45)+Q M^3( -30\omega^2+8)+Q^2 (96\omega^2
-36)\right]\nonumber\\
g_6(\omega)&=&(1-\omega^2) \omega (\frac{-1}{3696}) \left[M^6(7623 \omega^4
-6930 \omega^2+1155)+\right.\nonumber\\
&+&\left.Q M^3(25410  \omega^4+7546 -34188 \omega^2)+D M(9702  \omega^4 +27510-69580 \omega^2)+\right.\nonumber\\
&+&\left.Q^2(6930 \omega^4+52356\omega^2-20694)\right]\nonumber\\
g_7(\omega)&=&(1-\omega^2) \omega(\frac{-1}{490M^2}) \left[D Q M(-64680 \omega^4+54600 \omega^2-8400)+\right.\nonumber\\
&+&\left.Q^2 M^3(-19845\omega^4+7024 \omega^2+957)+Q M^6(
4410 \omega^4-3038 \omega^2+336)+\right.\nonumber\\
&+&\left. M^4 D(25725 \omega^4-13475 \omega^2-210)+Q^3(34776 \omega^4
-27468 \omega^2+4032)\right]\nonumber\\
g_8(\omega)&=&(1-\omega^2) \omega \frac{1}{1921920M} \left[Q^3(417624012 \omega^4-327877128 \omega^2+48074796)+\right.\nonumber\\
&+&\left.D Q M(-20270250 \omega^6-693406350 \omega^4+595505850 \omega^2-92457090)+\right.\nonumber\\
&+&\left.Q M^6(-58558500 \omega^6+100720620\omega^4+4832100-45945900 \omega^2)+\right.\nonumber\\
&+&\left.M^9(
-10735725 \omega^6+15030015 \omega^4-5780775 \omega^2+525525)+\right.\nonumber\\
&+&\left.M^4D(-37837800 \omega^6+226746520\omega^4-137167800 \omega^2+12251400)+\right.\nonumber\\
&+&\left.M^3Q^2(-49549500 \omega^6-83972460 \omega^4+63989772\omega^2-4272228)
\right]
\label{A2}
\end{eqnarray}

The following expressions show the relation of Weyl coordinates $\{R,\omega\}$ in terms of  the MSA coordinates $\{r,y\}$ up to order $O(\hat\lambda^9)$:

\begin{eqnarray}
\omega&=&\omega(r,y)=y+\frac 12 y (1 -y^2) \hat\lambda^2+\left( (1-y^2)+q (1-y^2)\right)y \hat\lambda^3+\nonumber\\
&+&\left[(-\frac94  y^2+\frac{15}{8} +\frac38  y^4)+q(-\frac54  y^4 -\frac{5}{14}  y^2 +\frac{45}{28})\right] y\hat\lambda^4+\nonumber\\
&+&\left[(-5 y^2 +\frac 72 +\frac32 y^4)+q (-\frac{22}{35} y^2 +\frac{92}{35}-2  y^4)+d(15 y^2-\frac92 -\frac{21}{2} y^4)+\right.\nonumber\\
&+&\left.q^2 (-\frac{66}{5} y^2+\frac{18}{5}+\frac{48}{5} y^4)\right] y\hat\lambda^5+\nonumber\\
&+&
\left[(\frac{75}{16}  y^4-\frac{5}{16}  y^6+\frac{105}{16}   -\frac{175}{16}  y^2)+q (-\frac{39}{8} y^4+\frac{15}{8} y^6+\frac{715}{168}-\frac{211}{168} y^2)+\right.\nonumber\\
&+&\left. q^2(\frac{9179}{616}+\frac{26617}{616} y^4-\frac{15}{8} y^6-\frac{34641}{616} y^2)+\right.\nonumber\\
&+&\left.d(-\frac{1325}{88} -\frac{8197}{264}  y^4-\frac{21}{8}  y^6+\frac{12865}{264}  y^2)\right] y\hat\lambda^6+\nonumber\\
&+&
\left[(-\frac{15}{8} y^6 +\frac{105}{8}  y^4-\frac{189}{8}   y^2+\frac{99}{8} )+ q(\frac{1893}{280} +\frac{57}{8} y^6-\frac{463}{40}  y^4-\frac{647}{280}  y^2)+\right.\nonumber\\
&+&\left. q^2(\frac{412336}{2695} y^4-\frac{929517}{5390} y^2+\frac{116833}{2695}-\frac{239}{10}  y^6) +\right.\nonumber\\
&+&\left. q^3(\frac{4446}{35} y^4-\frac{450}{7} y^2+\frac{288}{35}-\frac{2484}{35} y^6)+\right.\nonumber\\
&+&\left. d(-\frac{9981}{88} y^4+\frac{86511}{616}   y^2-\frac{24729}{616} +\frac{105}{8} y^6)+\right.\nonumber\\
&+&\left. q d(-\frac{1704}{7} y^4+\frac{900}{7} y^2-\frac{120}{7} +132 y^6)\right]y \hat\lambda^7+\nonumber\\
&+&
\left[(\frac{35}{128}  y^8+\frac{2205}{64} y^4 +\frac{3003}{128} -\frac{1617}{32}   y^2-\frac{245}{32} y^6)+\right.\nonumber\\
 &+&\left. q(-\frac{15119}{560} y^4+\frac{54881}{5280}-\frac{32327}{9240}   y^2+\frac{1257}{56} y^6-\frac{75}{32} y^8)+\right.\nonumber\\
 &+&\left. q^2(\frac{165}{32} y^8+\frac{272710509}{560560} y^4-\frac{416569}{3080} y^6-\frac{130608551}{280280} y^2+\frac{122863527}{1121120})+\right.\nonumber\\
 &+&\left. q^3(-\frac{1144047}{2464} y^6+\frac{26397765}{32032} y^4-\frac{13183965}{32032} y^2+\frac{21543}{416})+ \right.\nonumber\\
 &+&\left. d(\frac{938}{11} y^6-\frac{1241945}{3432} y^4+\frac{316955}{858} y^2-\frac{221363}{2288}+\frac{63}{16}  y^8 )+\right.\nonumber\\
 &+&\left. q d(\frac{273111}{352} y^6-\frac{22394167}{16016} y^4+\frac{3333245}{4576} y^2-\frac{556299}{5824}-\frac{675}{64} y^8) \right] y\hat\lambda^8
\label{A3}
\end{eqnarray}

\begin{eqnarray}
R&=&R(r,y)= r \left[1-\hat\lambda-\frac 12(1-y^2) \hat\lambda^2+\left(-\frac 12 (1-y^2)-q\frac 12 (1-3 y^2)\right) \hat\lambda^3+\right.\nonumber\\
&+&\left.\left((-\frac58 +\frac34 y^2-\frac18 y^4)+q( -\frac{9}{28} +\frac{3}{14} y^2 +\frac54 y^4)\right) \right.\hat\lambda^4+\nonumber\\
&+&\left.\left((-\frac38 y^4 -\frac78+\frac54 y^2)+q(2 y^4 -\frac{5}{14} y^2-\frac{3}{14})+q^2(9 y^2-12 y^4-1)+\right.\right.\nonumber\\
&+&\left.\left.d(-\frac{45}{4} y^2 +\frac{105}{8} y^4 +\frac98 )\right)\hat\lambda^5+\right.\nonumber\\
&+&\left.\left((\frac{21}{16}-\frac{15}{16} y^4+\frac{35}{16} y^2+\frac{1}{16}y^6)+q(-\frac{1}{168} +\frac{219}{56} y^4-\frac{11}{8} y^2-\frac{5}{8} y^6)+\right.\right.\nonumber\\
&+&\left.\left.q^2( -\frac{1525}{616}-\frac{21099}{616} y^4+\frac{15525}{616} y^2+\frac{15}{8} y^6)+\right.\right.\nonumber\\
&+&\left.\left.d(\frac{205}{88}-\frac{1995}{88} y^2 +\frac{21}{8} y^6 +\frac{2135}{88}y^4)\right) \hat\lambda^6+\right.\nonumber\\
&+&\left.\left((-\frac{33}{16}+\frac{5}{16} y^6-\frac{35}{16}  y^4+\frac{63}{16} y^2)+q(\frac{155}{336}+\frac{4363}{560} y^4 -\frac{1973}{560} y^2-\frac{33}{16} y^6)+\right.\right.\nonumber\\
&+&\left.\left.q^2(-\frac{1651}{308}-\frac{135637}{1540}y^4 +\frac{91493}{1540} y^2+\frac{287}{20} y^6)+\right.\right.\nonumber\\
&+&\left.\left.d(\frac{829}{
176}+\frac{10735}{176} y^4 -\frac{8631}{176}y^2-\frac{63}{16} y^6)+q d(3 +195 y^4-154 y^6-60 y^2)+\right.\right.\nonumber\\
&+&\left.\left. q^3(-\frac{3}{2}+\frac{414}{5} y^6+\frac{144}{5} y^2-\frac{981}{10} y^4)\right) \right.\hat\lambda^7+\nonumber\\
&+&\left.\left((-\frac{429}{128} -\frac{5}{128} y^8+\frac{35}{32} y^6 -\frac{315}{64} y^4+\frac{231}{32}  y^2)+\right.\right.\nonumber\\
&+&\left.\left.q d(\frac{722587}{64064}-\frac{113533}{176} y^6 +\frac{675}{64} y^8 +\frac{25975585}{32032} y^4 -\frac{360285}{1456}y^2)+\right.\right.\nonumber\\
&+&\left.\left.q^2(-\frac{661565}{61152}-\frac{120202913}{560560}y^4-\frac{55}{32} y^8+\frac{45937}{770} y^6+\frac{18225373}{140140} y^2)+\right.\right.\nonumber\\
&+&\left.\left.d(\frac{506395}{3432} y^4 +\frac{20849}{2288} -\frac{1897}{66} y^6-\frac{21}{16} y^8 -\frac{14745}{143}y^2 )+\right.\right.\nonumber\\
&+&\left.\left.q(\frac{1571}{1056}-\frac{159}{28} y^6 +\frac{15}{32} y^8 -\frac{2689}{330}y^2 +\frac{26459}{1680} y^4)+\right.\right.\nonumber\\
&+&\left.\left.q^3(-\frac{149437}{24024} +\frac{1206969}{3080} y^6 +\frac{71211}{520} y^2 -\frac{18927009}{40040} y^4 )\right) \hat\lambda^8\right]
\label{A4}
\end{eqnarray}

\section{Acknowledgments}
This  work  was partially supported by the Spanish  Ministry of Education and Science
 under Research Project No. FIS 2009-07238, and the Consejer\'\i a de Educaci\'on of the Junta de Castilla y
Le\'on under the Research Project Grupo de Excelencia GR234. We also want to thank the support of the Fundaci\'on Samuel Sol\'orzano Barruso (Universidad de Salamanca) with the project No.FS/8-2009.

\end{document}